\documentclass[10pt,english,envcountsame,latin9,utf8]{article}
\usepackage[T1]{fontenc}
\usepackage[latin9]{inputenc}
\usepackage{geometry}
\usepackage{verbatim}
\usepackage{mathtools}
\usepackage{amsmath}
\usepackage{amssymb}
\usepackage{graphicx}
\usepackage{esint}

\makeatletter
\usepackage{xcolor}
\usepackage{hyperref}
\hypersetup{final, colorlinks=true, linkcolor=black, urlcolor=black, citecolor=black, linkbordercolor=black, pdfborderstyle={/S/U/W 1}}

\usepackage[absolute,overlay]{textpos}
\usepackage[numbers]{natbib} 

\usepackage{listings}
\usepackage{blkarray}

\usepackage{tikz}
\usetikzlibrary{positioning,arrows,calc,fit,automata}
\usepackage{xcolor}

\usepackage{pgfplots}
\DeclareUnicodeCharacter{2212}{}
\usepgfplotslibrary{groupplots,dateplot}
\usetikzlibrary{patterns,shapes.arrows}
\pgfplotsset{compat=newest}

\usepackage{nccmath}
\usepackage[skins]{tcolorbox}
\usepackage{authblk}
\date{}
\makeatother

\usepackage{babel}
\begin{document}
\onecolumn

\title{Friction Interventions to Curb the Spread of Misinformation on Social Media}

\author[1]{Laura Jahn}
\author[1]{Rasmus K. Rendsvig}
\author[2]{Alessandro Flammini}
\author[2]{Filippo Menczer}
\author[1]{Vincent F. Hendricks}
\affil[1]{Center for Information and Bubble Studies,
University of Copenhagen, Denmark}
\affil[2]{Observatory on Social Media, Indiana University, Bloomington, USA}
\maketitle

\begin{abstract}
\noindent Social media has enabled the spread of information at unprecedented speeds and scales, and with it the proliferation of high-engagement, low-quality content. 
\emph{Friction}---behavioral design measures that make the sharing of content more cumbersome---might be a way to raise the quality of what is spread online. 
Here, we study the effects of friction 
with and without quality-recognition learning. 
Experiments from an agent-based model suggest that friction alone decreases the number of posts without improving their quality. 
A small amount of friction combined with learning, however, increases the average quality of posts significantly. 
Based on this preliminary evidence, we propose a friction intervention with a learning component about the platform's community standards, to be tested via a field experiment. 
The proposed intervention would have minimal effects on engagement and may easily be deployed at scale.

\end{abstract}

\bigskip{}

The spread of misinformation online has been recognized as a global, societal threat to democracy, eroding trust in mainstream news sources, authorities, experts, and other socio-political institutions \cite{HendricksMehlsen2022,CRA-agenda-2020,Shao_Menczer_NatComm2018,WEF2018}.
Social media communication platforms have enabled the sharing of information at unprecedented speeds and scales, and with it the proliferation of not only misinformation, but also other low-quality and harmful content such as hate speech, cyberbullying, and malware \cite{CRA-agenda-2020,Hendricks2021}. 
Large social media platforms amplify the so-called attention economy, where abundant information competes for scarce attention \cite{HendricksMehlsen2022,simon1971informationrich,ciampaglia2015attentionecon}. 
Through this competition, one would hope for accurate information to emerge from the interactions
among many users by combining \emph{independent} opinions according to the \emph{wisdom of crowds} \cite{surowiecki2005wisdom}. 
Alas, scholars have demonstrated that engaging yet false content gets shared more and travels faster \cite{vosoughi2018spread}.
The focal question of this paper is whether adding a bit of \emph{friction} to the sharing process might mitigate the spread of low-quality and malicious content.

Socio-cognitive biases and algorithmic sorting both contribute to the spread of high-engagement content over high-quality content. 
Strategies such as accepting new information if it comes from multiple sources \cite{centola2007complex} or through posts that have been shared many times fail because the aggregate opinions to which we are exposed online are not necessarily independent \cite{lorenz2011social}. 
This is boosted by confirmation bias \cite{nickerson1998confirmation}, a disconnect between what users deem accurate and what they deem shareable \cite{pennycook2022accuracy,Pennycook2021}, and automatic habits to share the most engaging content \cite{Ceylan23_habitual}. 
The illusory truth effect \cite{lacassagne2022illusionary} further increases vulnerability to misinformation in social media by increasing the perceived truth value of low-quality content through repetition \cite{hills2019dark, fazio2015illusionaryeffect}.

Algorithmic biases in the content sorting algorithms of social media platforms also prioritize high engagement  \cite{nikolov2019quantifying}, increasing the exposure of low-quality content in user news feeds \cite{ciampaglia2018algorithmic}. 
One-click reactions, such as ``Like'' or ``Share,'' are a driving mechanism behind algorithmic sorting since they are easy to use and quick to influence the popularity of posts. 
Users can select a reaction to a post from a short list, with their aggregate choices typically presented as a popularity metric beneath the post. 
These reactions steer user attention. 
For example, a high retweet count is likely to be perceived as a crowd-sourced trust signal \cite{surowiecki2005wisdom,Metaxas2015}, possibly contributing to the content's virality \cite{dutta2018retweet} irrespective of its quality \cite{avram2020exposure}. 
One corollary of these dynamics is an incentive for influence operations based on inauthentic behaviors, such as coordinated liking \cite{Torres-Lugo_Likes_Manip_deletions, nizzoli2021coordinated, Orabi2020,Goerzen2019, Ferrara2017, Ferrara:2016, Takacs2019, Pacheco2021Coordinated, duan2022botsIndiana}.

Scholars have called for ways to promote the Internet's potential to strengthen rather than diminish democratic virtues and public debate \cite{Lazer2018} and to leverage the economics of information for protection rather than for misguidance \cite{MenczerHills2020SciAm}.
A relatively recent idea is to improve the quality of what is shared online by introducing \emph{friction} on social media. The hope is to curb the spread of harmful content and misinformation by making it more difficult to share or like content online \cite{HendricksMehlsen2022,lou2019manipulating,MenczerHills2020SciAm}.

In the context of online interactions, \emph{behavioral friction}, in general, denotes ``any unnecessary retardation of a process that delays the user accomplishing a desired action'' \cite{Tomalin2022friction}. 
The more friction, the lower the chances that the user will complete the action. 
While reducing friction is generally deemed desirable in user interface design, some protective friction, like CAPTCHAs, may be useful \cite{Tomalin2022friction,InfoandDemocracy_policy2018,goodman2020friction}.
Friction added to otherwise one-click sharing and liking will make the spread of both harmful and benign content more cumbersome and time-consuming. 
Friction is thought to prompt a more deliberate approach to sharing or liking content. 
Examples include exposing users to a contextual label \cite{InfoandDemocracy_policy2018}, impeding the completion of an action with a prompt asking the user to reflect \cite{pennycook2022accuracy}, exacting micro-payments, or requiring users to spend mental resources through micro-exams such as quizzes and puzzles \cite{Shao_Menczer_NatComm2018, Hendricks2021}.
Such friction strategies promise to deliver socio-political benefits by supporting cognitive autonomy while increasing the cognitive burden of sharing low-quality content \cite{goodman2020friction}.


Borrowing terminology from Tomalin \cite{Tomalin2022friction}, the type of friction of interest in this paper is \emph{non-elective} for users---users cannot control exposure.  
This characteristic is also present in \emph{sludges} and \emph{dark patterns.} 
Sludges generally refer to excessive friction, bad almost by definition and with a clear negative valence (e.g., bureaucratic form-filling). 
Dark patterns coerce, steer, or deceive people into making unintended and potentially harmful choices \cite{SUNSTEIN2020}.
The types of friction we consider in this paper aim to obtain social benefits. They are easily distinguishable from sludges and dark patterns as they are \emph{overt}, neither \emph{deceptive} nor \emph{accidental}, but \emph{intended}, \emph{protective}, and \emph{non-commercial} \cite{Tomalin2022friction}.
Friction strategies that share these characteristics may be \emph{impeding} or \emph{distracting}. For example, friction can impede an action by letting users complete it only after a micro-exam is passed.
Certain nudges provide non-impeding friction; deliberation-promoting nudges, for example, are distracting but leave all options available to the user \cite{SUNSTEIN2020}.

Recent research suggests that non-elective, overt, intended, protective friction is a promising tool to boost the accuracy and quality of information shared online. Adding as little friction as having users pause to think before sharing may prevent misinformation proliferation on social media: 
in a set of online experiments, participants who were asked to explain why a headline was true or false were less likely to share false information compared to control participants \cite{Fazio2020}.
In an effort to nudge users to consciously reflect on tweet content, Bhuiyan et al. \cite{2018Bhuiyan_FeedReflectgameintervention} developed a browser extension that introduced a distracting emphasis on high-quality content and greyed out posts from low-quality sources. This raised the accuracy of tweet credibility assessments.
Pennycook et al. \cite{Pennycook2021,pennycook2022accuracy} see potential in reminding users of accuracy. They prompted experiment participants to rate the accuracy of a news headline before scrolling through an artificial social media news feed. Participants subsequently shared higher-quality content than a control group.
The authors suggest to translate their findings into attention-based interventions subtly reminding users of accuracy to slow down the sharing of low-quality content online. Similarly, checking for accuracy assisted the fight against the illusory truth effect. This was demonstrated in a set of experiments that prompted participants to behave like fact checkers by asking them for initial truth ratings at first exposure \cite{brashier2020initialaccuracy}.

On Facebook, priming critical thinking made users less prone to trusting, liking, and sharing fake news about climate change \cite{LUTZKE2019}.
Reminding users of critical thinking was accomplished through considerations about news evaluation guidelines, using questions to help identify fake news as articulated in the Facebook Help Center (e.g., ``Does the information in the post seem believable?''). 
Time pressure has further been shown to negatively influence the ability to distinguish true and false headlines \cite{sultan2022timepressure}. Friction---as a means to reduce time pressure---may actively improve the discrimination of accurate and false information. 
Lastly, friction has been deemed beneficial when applied to mass-sharing. Model-based work by Jackson et al. \cite{jackson2022learning} shows that caps on depth (how many times messages can be forwarded) or breadth (the number of others to whom messages can be forwarded) improve 
the ratio of true to false messages, assuming messages mutate at every instance of re-sharing (deliberately or inadvertently).

Most interventions by social media platforms to date do not impose restrictions on sharing \cite{CarniegeIndexInterventions}. They tend to use redirection (suggesting content from authoritative sources such as the WHO during the COVID-19 pandemic) and content labeling (exposing users to additional context), thus preserving user choice and autonomy.   More harsh but less common approaches include removing or down-ranking such content, banning users who spread it, or decreasing their reach. 
Various platforms have adopted friction interventions.
Twitter implemented protective, non-elective friction that distracts or impedes. The platform introduced caps on automated tweeting \cite{MenczerHills2020SciAm} and a distracting label to pause users about to share state-affiliated media URLs \cite{Twitter_StateAff}. With limited success, they tested replacing retweets with ``quote tweets,'' requiring users to comment before they could share a post \cite{abp_retweet_back}. 
Twitter also conducted a promising randomized controlled trial to curb offensive behavior, where users were asked to review replies in which harmful language was detected \cite{katsaros2021reconsidering}.
%
Facebook has established policies to provide context labels from fact checkers. The platform reduces the distribution of, and engagement with, misinformation from repeated offenders by reducing the reach and visibility of their posts. 
While this intervention leads to a decrease in engagement with the offender in the short term, this can be compensated by an increase in the offender's posts and followers. Furthermore, the limitation in reach can be reversed by deleting flagged posts \cite{thero2022facebook}. 
%
WhatsApp has taken first steps to counter the virality of misinformation by limiting the forwarding of messages to at most five contacts simultaneously \cite{DeFreitasMelo2020}.
%
Instagram has introduced a distractive but non-impeding anti-bullying label that prompts users to pause by asking them ``Are you sure you want to post this?'' to curb abuse on the platform \cite{Lee2019BBCInsta}.

Unfortunately, reporting by social media platforms on the friction interventions deployed to date lacks transparency around the testing and implementation process, making it difficult for researchers to study the different countermeasures. 
In addition, the effectiveness of content labeling has been challenged by research findings \cite{brashier2020initialaccuracy}: while the common advice for dealing with fake news is to consider the source, people often struggle to remember sources \cite{henkel2011reading}. 
Tagging only some false news stories as ``false'' may boost the perceived accuracy of inaccurate but untagged stories due to an implied truth effect \cite{pennycook2020implied}.

In this paper, we aim to systematically explore how friction may positively affect information quality in a social media environment. 
We study the effects of friction prompts with and without quality-recognition learning components through an agent-based model (ABM). 
Our experiments suggest that friction alone decreases the number of posts without improving their average quality. On the other hand, a small amount of friction combined with learning increases the average quality of posts significantly. 
Inspired by this preliminary evidence, we propose a friction intervention where learning is leveraged through quizzes about a platform's community standards.
We map the key ingredients of a field experiment to test the idea; platforms could facilitate larger studies to test scalable friction strategies.

\section*{Results}

\subsection*{Model}

To study the impact of friction, we carry out a proof-of-concept study. We add mechanisms for friction and learning to \emph{SimSoM}, a minimal open-source agent-based model of information sharing in social media \cite{lou2019manipulating, SimSom23}.

In the augmented ABM, \emph{posts} are interpreted as pieces of information, such as images, links, hashtags, or phrases. These may be created or shared by \emph{agents}, and appear on the \emph{news feeds} of agents. Each agent's news feed consists of a bounded number of posts, all shared by agents they \emph{follow}. The bounded news feed models limited individual attention, which gives rise to heavy-tailed distributions of post popularity and lifetime consistent with empirical data \cite{weng2012competition}. 

The ABM runs in discrete time. 
At each time step, some agents are \emph{activated}. Each chooses to either \emph{introduce} a new post into the network, or \emph{share} an existing post from its news feed. The new or re-shared post then appears on the news feeds of the agent's followers.
A time step is interpreted as a social media session where only a sample of all users are online simultaneously.

Posts may vary in \emph{quality} and in how \emph{engaging} they are. 
Quality models an objective desirable property, such as accuracy or relevance of posts. Engagement models the quality of a post as \emph{perceived} by agents. 
The quality and engagement of a post are sampled such that low-quality posts are more likely than high-quality posts, and low-engagement posts are more likely than high-engagement posts. 
Quality and engagement are sampled independently to reflect that high quality and high engagement  do not necessarily coincide \cite{vosoughi2018spread}. 
While the ABM does not encode agent types, low-quality posts may be thought to stem from a variety of accounts, such as authentic human users, social bots, cyborgs, or algorithmic amplifiers \cite{yan2022botlandscape},  broadly understood.

\begin{figure}
\includegraphics[scale=0.33]{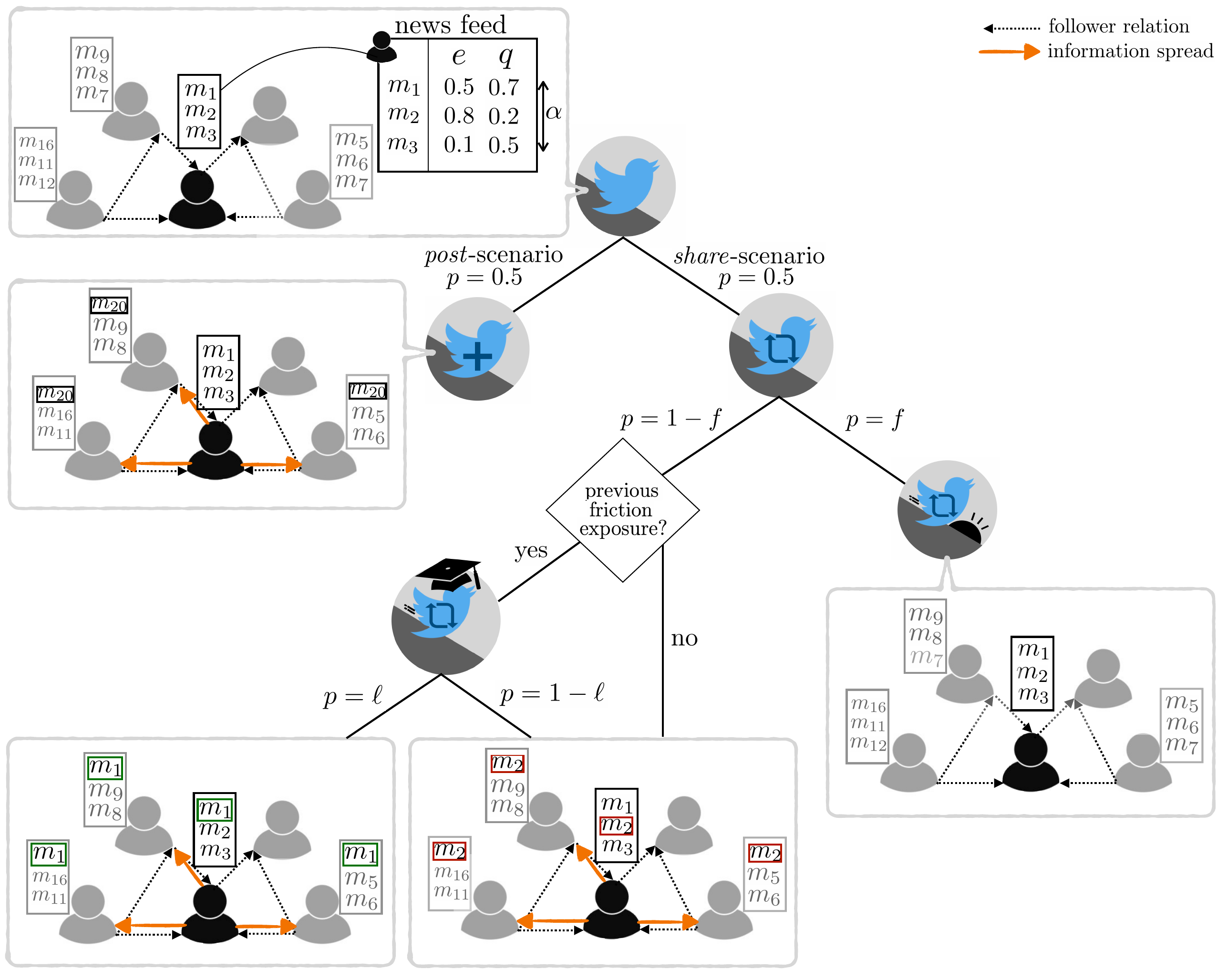}\caption{\label{fig:Sim}
\textbf{Information diffusion process.}
    Each node has a news feed of size $\alpha$, containing messages recently posted or re-shared by friends. 
    The follower relation is illustrated by dotted arrows pointing from an agent to their friends.
    Information travels from agents to their followers, along the orange arrows (in the opposite direction of the follow relations).
    At each time step, a subset of agents act (here, the central black node). 
    With probability 0.5, an acting agent posts a new message (here, $m_{20}$), else they re-share a message from their news feed. The new message appears at the top of the followers' news feeds and the existing messages are shifted down. If $\alpha$ messages were already present in a feed, the oldest one is discarded. 
    If the agent shares, with probability $f$, they are exposed to friction and prevented from sharing. Else, with probability $1-f$, they scan their feed and share a post in their feed selected with probability proportional to the post's engagement (here, $m_{2}$)---unless  the agent has been exposed to friction earlier and learned (with probability $\ell$): in that case, the agent instead selects the post to share with probability proportional to the post's quality (here, $m_{1}$).}
\end{figure}

In our model, friction is restricted to the stage when agents re-share something posted by their friends (Fig.~\ref{fig:Sim}). In this \emph{share} scenario, agents may face a friction prompt. 
The intuition is that friction triggers agents to pause, potentially impeding their sharing activity. 
Agents may resume re-sharing the chosen post after having spent mental resources, or passed a quiz.
On the other hand, agents do not resume sharing the chosen post if they either reconsider or fail to comply with the quiz. 
The probability that agents in the \emph{share} scenario are \emph{exposed} to friction and either \emph{reconsider} or \emph{fail} to comply is captured by the parameter $f \in [0,1]$ (see Fig.~\ref{fig:Sim}).  
The simulation records that an agent has been exposed to the friction prompt.

Agents may learn through exposure to a friction prompt, e.g., through deliberation-triggering nudges or educational quizzes that remind agents to pay attention to quality. 
Next time an agent is about to re-share a post, an agent who has learned no longer re-shares the most engaging post, but instead chooses a post to re-share based on quality. 
The parameter $\ell \in [0,1]$ controls the probability with which agents learn after previous friction exposure.
We call this type of learning \emph{quality-recognition learning}, drawing intuition from research both on priming effects and nudges \cite{weingarten2016priming_meta, pennycook2022accuracy, LUTZKE2019}, and on testing effects and retrieval practices shown to boost learning \cite{paul2015learn_testing, rowland2014testing_effect, endres2015testing_retrieval}. 
Learning through priming and nudging takes place without conscious guidance. In contrast, testing and retrieval of previously absorbed knowledge takes place consciously. 
Yet, both serve as learning events. Which mechanism takes precedence depends on the design of the friction prompt. Learning in the ABM and the model implementation is kept at a general level and allows for both intuitions, as the outcome---learning to recognize quality---does not change. The implemented friction strategy may thus fall into the \emph{headline-discernment paradigm} (discerning  true and false headlines and indicating willingness to share them)  and \emph{skill-adoption paradigm} (learning the skills and strategies required to evaluate information quality) of research on behavioral interventions \cite{kozyreva2022toolboxinterventions}. See Methods for more details on the SimSoM information diffusion, and the learning model and its parameters.
\bigskip{}

\subsection*{Information Quality and Discriminative Power}

To quantify the effect of friction and learning, our simulation tracks the changes in two metrics that capture desirable properties of an online social network:
the average quality of posts in the network and the system's capacity to discriminate information on the basis of its quality. At the end of each run, the \emph{popularity} of posts is measured by the number of times posts have been shared or re-shared.

Ideally, one wants the system to discriminate against low-quality posts by reaching a strong correlation between quality and popularity: 
the higher the quality of a post, the more widely it should be shared among agents. 
We capture this discriminative power by measuring Kendall's rank correlation~\cite{kendall1945treatment} between popularity and quality of posts. 

\begin{figure}
\begin{center}\includegraphics[scale=1]{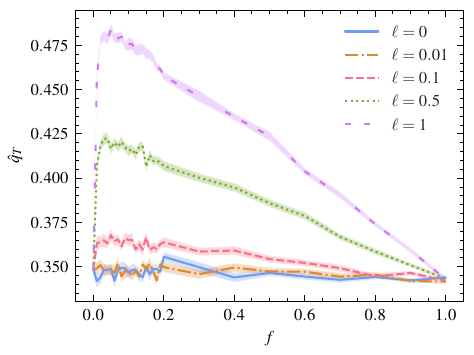}\end{center}
\caption{\label{fig:quality} Average post quality $\hat{q}_{T}$ as a function of friction probability $f$, for different probabilities of learning $\ell$. 
The subscript $T$ indicates that average quality is measured at convergence (see Methods). 
Shaded areas indicate standard errors.}
\end{figure}

\begin{figure}
\begin{center}\includegraphics[scale=1]{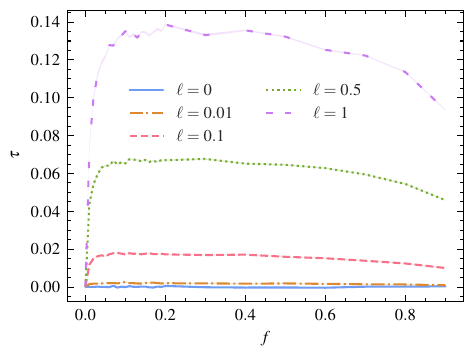}\end{center}
\caption{\label{fig:discriminativepower} Average discriminative power $\tau$ as a function of friction probability $f$, for different probabilities of learning $\ell$. Shaded areas indicating standard errors are too small to be visible.
The Kendall rank coefficient is not defined for $f=1$, as in this case nothing gets re-shared and the popularity ranking is all ties (see Methods).}
\end{figure}

To assess how friction and associated learning affect information diffusion of high- and low-quality posts in the networks, we plot average quality $\hat{q}_{T}$ in Fig.~\ref{fig:quality} and discriminative power $\tau$ in Fig.~\ref{fig:discriminativepower}. 

Fig.~\ref{fig:quality} summarizes how friction affects average quality: 
in the absence of learning, we do not observe an increase in quality when solely adding friction to the network. 
This is plausible, as quality and engagement values of posts are sampled independently, and friction is triggered independently of the quality or engagement of the post. 
The findings suggest a significant increase in average quality when $f=0.1$, (i.e., the probability for an agent being prevented from re-sharing is $0.5 \cdot 0.1 = 0.05$) as long as friction-exposed agents learn, even if only rarely (e.g., $\ell \geq 0.1$). 
Higher learning probabilities boost quality significantly. For $\ell=0.5$, $\hat{q}_{T} \approx 0.41$ and for $\ell=1$, we observe $\hat{q}_{T} \approx 0.47$---an increase by more than one third. 

Fig.~\ref{fig:quality} also reveals that increasing the probability of friction exposure to more than $f=0.1$ does not result in higher quality, no matter how well agents learn. 
These findings suggest that in the presence of learning, only a little friction provides the best conditions for agents to apply learned behavior, even when accounting for forgetting or making mistakes ($\ell<1$). 
The reason for quality to drop with high $f$ is that agents cannot benefit from newly won awareness of posts' quality when they are prevented from sharing. 
Therefore, combining rare friction prompts with quality-recognition learning is far more desirable than extensively restricting sharing.

There is a trade-off between average quality and diversity in posts. The maximum quality is obtained when the best post is featured on all agent news feeds, which kills diversity. 
As an artifact of friction preventing re-sharing, diversity increases with friction in the model. 
In the presence of both high quality and diversity, $\tau$ lets us assess how well the network can discriminate between high and low-quality (Fig.~\ref{fig:discriminativepower}). 
We again find that friction combined with learning significantly increases the system's discriminative power. 

If agents fail to learn, friction yields no improvements in discriminative power ($\tau \approx 0$). 
With induced friction levels as little as $f=0.1$ and associated learning $\ell=0.1$, we find higher correlation between post popularity and quality. Discriminative power is highest for higher $\ell$ and $f$ between $0.1$ and $0.2$: $\tau$ increases up to $0.139$ at $f=0.2$.
Increasing friction to more than $f=0.2$, does not further improve the network's discriminative power, as agents are more likely to be stopped from employing their learned awareness of quality.

\section*{Discussion}

Preliminary results from the agent-based model suggest that a small amount of friction combined with quality-recognition learning increases the average quality of posts significantly. 
Both the average quality and the network's power to discriminate between high- and low-quality posts increase with just a bit of friction even with far-from-perfect learning.
On the other hand, excessive friction, even with perfect learning, has a negative effect on average quality as agents are prevented from applying their learnings.
In light of these theoretical findings, let us discuss a concrete idea for a friction prompt and outline design avenues for a field experiment. 

A concrete friction prompt could quiz users about community standards to promote quality recognition before they can react to content. 
In an experimental environment, users would be faced with randomly assigned micro-exams (e.g., multiple choice) when they are about to share or like posts. 
The idea of impeding quizzes has most famously been established with CAPTCHA tests. 
Quizzes testing a reader's understanding of an article have also been successfully used in the comment section of a public Norwegian broadcaster \cite{NiemanLab_Norwayfriction}. 
By involving questions about the platform's governing community standards, the exams would help users learn about such norms \cite{Hendricks2021}. 
Lutzke et al. \cite{LUTZKE2019}  showed that the combination of reminding users of critical thinking and questions taken from community standards may help participants identify fake news. 
The display of community norms on Reddit increased compliance and prevented unruly and harassing conversations \cite{matias2019commstandardsreddit}.

The proposed study would measure whether engagement with low-quality posts drops when users are prompted with friction strategies incorporating quality-recognition learning based on community standards. 
Engagement with low-quality posts, measured by likes and shares, should be compared to engagement with high-quality content.
A possible definition of low-quality posts could be based on violations of community standards, such as those on misinformation. 
Such a field experiment may also contribute to research on testing effects \cite{paul2015learn_testing, rowland2014testing_effect, endres2015testing_retrieval}, nudges, and priming \cite{pennycook2022accuracy, LUTZKE2019}. 

In the proposed field experiment, participants will engage with posts as they would on their preferred platform. 
The friction intervention will take the form of a prompt during this process. 
Quizzes would randomly remind users of the platform's community standards (not only upon violating those, in contrast to Katsaros et al. \cite{katsaros2021reconsidering}). 
Quizzes may query the participants about, say, definitions, examples, and risks of violating misinformation or other community standards.
The design is intended to promote fluency in the community standards rather than just awareness of particular sanctioning clauses.
Both intervention and control groups should be set up to test for causality \cite{epstein2022sharingexperiment}, rather than exposing all study participants to the designed intervention as in Avram et al. \cite{avram2020exposure}. 
We also refer to Pennycook et al. \cite{pennycook2021practicalguide} for a comprehensive guide on behavioral experiments related to misinformation and fake news. 

Unfortunately, testing such friction strategies in a real environment poses difficulties \cite{misinfo_data}. 
Independent researchers do not have access to a platform such as Twitter to design and test real-time interventions in randomized controlled trials.
This approach is therefore neither feasible nor reproducible, except for researchers employed by social media companies. 
One may instead carry out the experiment in a system designed to emulate a social media environment
Available options include Amazon Mechanical Turk \cite{Pennycook2019,Fazio2020}, Volunteer Science \cite{VolunteerScience}, games such as Fakey \cite{Fakey2021}, 
or open-source software such as the Mock Social Media Tool \cite{MockSocialMediaTool}. 
Each environment yields different levels of ecological validity, and experiments may be informed by empirical data about user activity and social network structure guiding online information sharing \cite{DeFreitasMelo2020}.

A key design decision is what to show participants in their feeds.  
The higher the desired ecological validity, the more sizeable the task of labeling tweets to determine intervention effects. 
One extreme is to work solely with each user's own feed. 
While this yields the highest ecological validity, it comes with two downsides. First, one cannot be sure that users are exposed to misinformation at all, making the size of a sufficient data sample unknown. Second, all posts users have seen, or at least engaged with, must be labeled as misinformation or not. 
Labeling has either limited accuracy when derived from lists of low-credibility sources \cite{lin2022high,Shao_Menczer_NatComm2018}, or requires a lot of manual labor. 
Alternatively, one may work with synthetically curated feeds.
A set of posts could be (partially) curated by researchers \cite{sultan2022timepressure,pennycook2021practicalguide,Fazio2020,pennycook_mcphetres_zhang_lu_rand_2020,Wangmisinfotweets2021,SHAHI2021misinformcovid19twitter}. These posts could be shown to all participants, thus ensuring that all users are shown the same misinformation posts while restricting the set of posts that must be labeled. 
Exposing participants to injected low-credibility content may raise valid ethical concerns, creating a potential need for mental health support during the study \cite{pennycook2021practicalguide}. 
The experimental design will have to weigh the advantages and limitations of both real and synthetic feeds.

As a policy, the friction strategy proposed here is non-elective, overt, protective, non-commercial, and impeding. 
While impediment may imply intrusiveness, our results show that in the presence of learning, only a little friction provides the best conditions for agents to apply learned behavior. Therefore we believe that a successful implementation would minimize impediments. 
In contrast, the redirection and content labeling strategies that dominate platform interventions are merely distracting. This allows harmful content to continue to circulate and places a greater burden on users to manage threats from such content themselves. Users who ignore labels or do not follow redirections remain vulnerable. 

An important advantage of friction prompts is that they circumvent the task of identifying bad actors, which is resource-expensive and difficult \cite{openletter,Martini2021,CRA-agenda-2020,misinfo_data,Shao_Menczer_NatComm2018,Yang2022Botometer101}. 
In contrast to other friction-based strategies \cite{katsaros2021reconsidering, 2018Bhuiyan_FeedReflectgameintervention, thero2022facebook}, our approach does not rely on classification and labeling of content as high- or low-quality; one-click reactions can be interpreted as supporting the original post without invoking NLP techniques or human annotation to determine the tone of comments/replies. 
Friction prompts presents an unbiased, scalable strategy \cite{kozyreva2022toolboxinterventions} that actively adds costs for inauthentic actors. 

We conclude with a call to further explore, test, and experiment with friction that educates users on community standards and triggers quality-recognition learning on social media platforms. 
On the one hand, greater familiarity with community standards will ease the burden of enforcing the standards. On the other, users who are well informed about community standards may encourage platforms to enforce their content moderation rules transparently and consistently.
Hopefully, learning about and enforcing community standards will stimulate public debate pertaining to online public spaces and democracy, freedom of expression, privacy, and user rights---principles often affirmed by tech giants themselves. 
As also required by the recent Digital Services Act (DSA) of the European Union \cite{DSA}, we encourage any platform undertaking such an intervention to share data and tools with researchers so they may aid in assessing the intervention's efficacy.

\section*{Methods}

The \emph{SimSoM} model \cite{lou2019manipulating, SimSom23} is a simple agent-based model of information sharing on social media. We amend the model by adding mechanisms for friction and learning. Next we describe network assumptions, the modeling of information diffusion, friction and learning, and the measurement of descriptive metrics together with simulation run details. 

\subsection*{Networks}

\emph{Networks} in the ABM are directed graphs with vertices representing agents and edges representing follower relations. If there is an edge from agent $i$ to agent $j$, we say that $i$ \emph{follows} $j$.
Agent $i$ then pays attention to content shared by $j$.
Each network consists of $N=1{,}000$ agents, and structurally mimics online social networks. To capture the characteristic presence of hubs, we construct networks using a directed variant of the {Barab\'asi-Albert} preferential attachment mechanism \cite{barabasi1999emergence}. 
A network is initialized with $m=3$ fully connected nodes and is grown by attaching new nodes each with $m$ outgoing edges that are preferentially attached to existing nodes with high in-degree. 
Put differently, agents with many followers (high in-degree) attract more followers. 
This results in scale-free properties and few, highly influential agents, i.e., influencers, mirroring the well-documented Matthews effect in action as it relates to follow relationships \cite{Merton1968,Perc2014}. 
We also wish to capture the characteristic presence of clustering (directed triadic closure \cite{Weng2013_triadicclosure}). To this end, we add edges by randomly sampling the friend of a friend of a target node and have the target node follow the sampled agent.  This has the effect of generating directed triads. We do this until the undirected clustering coefficient reaches $0.29$, as measured in a large sample from the empirical Twitter follower network \cite{nikolov2021followernetwork}. 
Networks are generated before a simulation run starts and do not change during a run.

\subsection*{Information Diffusion} 
\label{subsec:Network-and-diffusion}

At each time step in the agent-based model, $N$ agents are randomly selected to act in sequence. With probability $p=0.5$, a sampled agent $i$ posts a new message, otherwise $i$ re-shares a post from their news feed. Call the first the \emph{post} scenario and the latter the \emph{share} scenario (see Fig.~\ref{fig:Sim}). 
This modeling choice reflects the approximate average ratio of original tweets (vs. retweets) per agent, as measured in a large-scale sample of English-language tweets \cite{alshaabi2021growing}.
The new or re-shared post is added to the news feeds of $i$'s followers.  

Each agent's feed contains the $\alpha$ most recent messages posted or re-shared by those they follow, i.e. their friends; if a feed exceeds $\alpha$ posts, the oldest is discarded. 
Although social media platforms do not usually sort posts in strict reverse chronological order, this is a reasonable simplifying assumption because all platforms give high priority to recent posts. 
The parameter $\alpha$ models the number of posts viewed in a news feed during a session, and represents the finite attention of the agents.

Limited attention has been explored in previous work \cite{weng2012competition} and measured empirically on a social media mobile app as the number of times that a user scrolls at least 500 pixels through their feed and then stops for at least one second during an active session (idle time less than 30 minutes) \cite{qiu2017limited}. Following this measure, we adopt $\alpha=15$. 

Posts differ in quality and in how engaging they are. 
Both the quality $q$ and the engagement $e$ of a post are defined in the unit interval.
We independently draw a post's quality and engagement from the normalized probability density function $P(x)=\frac{(1-x)}{\intop_{0}^{1}(1-x)dx}=2(1-x)$ with $x=\{q,e\}$.\footnote{The sampling is implemented using inverse transform sampling, given the cumulative distribution function $C(x)=\intop_{0}^{1}P(x)dx=\intop_{0}^{1}2(1-x)dx$.} 
This simple linear distribution reflects the intuition that high-quality and high-engagement information are more rare. 

Initially, an agent cannot discern the quality of posts in their feed, but only the perceived quality, that is the engagement $e$. 
We assume that the probability that an agent re-shares a post is proportional to the post's engagement.
More explicitly, let $M_{i}$ be the feed of $i$ $(|M_{i}|=\alpha).$ 
The probability of post $m \in M_{i}$ being selected is $P(m)=e(m)/\sum_{j \in M_{i}} e(j)$ where $e(m)$ is the engagement of post $m$. This models cognitive bias: high-engagement posts will appear at a higher rate in agents' news feeds, further improving the odds of getting spread more. 
While the engagement $e$ of each post is fixed and does not change when a post gets shared more often, 
a news feed may contain duplicates of the same post: this happens, e.g., if an agent follows two others that share the same post. This further increases the chances that the duplicated message is re-shared, implicitly modeling an algorithmic bias that amplifies popular messages. 

\subsection*{Friction and Learning}

Exposure to friction happens in the \emph{share} scenario. The friction parameter $f \in [0,1]$ captures agents refraining from sharing as they either \emph{reconsider} or \emph{fail} to comply with a quiz.
Therefore the probability that an agent is prevented from re-sharing due to friction is $0.5 \cdot f$. 
Agents who have previously been exposed to friction prompts learn to discriminate between engagement and quality. Without prior exposure to friction, no learning happens. Formally, with probability $\ell$, an agent $i$ previously exposed to a friction prompt selects a message $m \in M_{i}$ with probability $P(m)=q(m)/\sum_{j\in M_{i}}q(j)$ where $q(m)$ is the quality of the post $m$. 
If $i$'s feed is only populated with posts having $q=0$, $i$ refrains from sharing any post and does not act at all. 
Through the parameter $\ell$, the model does not assume that agents are always able to apply what they have learned from one-time exposure to friction; they may still make mistakes or forget. 

\subsection*{Descriptive Metrics and Simulations Runs}

The measurements of average quality and discriminative power are averaged across simulation runs. The overall quality of an information ecosystem is given by average quality $\hat{q}_{T}$, measured at the end of a run.  
Each simulation run halts at the first time $T$ for which the exponential moving average quality of posts in the network's feeds stabilizes. 
At each time step $t$, we measure the average quality across all posts visible through the feeds of all the agents as $q=\frac{1}{aN}\sum_{i=1}^{N}\sum_{m\in M_{i}}q_{im}$, where $q_{im}$ is the quality of the $m$th post of agent $i$'s feed. 
We compute the exponential moving average $\hat{q}_{_{t}}=\rho\cdot\hat{q}_{_{t-1}}+(1-\rho)q_{_{t}}$, with smoothing factor $\rho=0.99$.  
We define stabilization by concluding the run at the first time $T$ for which $\mid\hat{q}_{_{T}}-\hat{q}_{_{T-1}}\mid<\varepsilon$ with $\varepsilon=10^{-5}$. Robustness tests showed no systematic changes in the trend of average quality with smaller $\rho$ nor smaller $\varepsilon$, but only longer runtimes. 
We record each post's quality and popularity (how often a post is shared or re-shared) at time $T$.

We average $\hat{q}_{T}$ and $\tau$ across a set of five sampled networks, for each parameter combination of induced friction $f$ and learning capability $\ell$. Our parameter combinations are all values for $f$ and $\ell$ in increments of $0.01$ until $0.2$ and in increments of $0.1$ after. Each of these combinations is run $10$ times per network. We thus analyze $50$ runs for each of 813 combinations of $f$~and~$\ell$.

\section*{Code and Data Availability}

The code for the agent-based model and its analysis is freely accessible on the public GitHub repository \emph{Friction-Social-Media-Model}.\footnote{See \url{https://github.com/LJ-9/Friction-Social-Media-Model}.} 
All data and results are reproducible.

\bibliographystyle{IEEEtranN} 

\end{document}